\documentclass[%
 aip,
 amsmath,amssymb,
 reprint,
]{revtex4-1}

\usepackage{graphicx}
\usepackage{dcolumn}
\usepackage{amsmath,amssymb}
\usepackage{bm} 
\usepackage{mathrsfs}
\usepackage[hidelinks]{hyperref}
\usepackage[title]{appendix}
\usepackage{comment}
\usepackage{color}
\usepackage[dvipsnames]{xcolor} 
\usepackage{outlines} 
\usepackage[utf8]{inputenc}
\usepackage[T1]{fontenc}
\usepackage{verbatim}
\usepackage{mathptmx}
\usepackage{etoolbox}
\usepackage{subcaption}
\usepackage{hyperref}
\hypersetup{
    colorlinks=true,
    linkcolor=blue,
    citecolor=blue,
    filecolor=magenta,      
    urlcolor=blue,
    pdfpagemode=FullScreen,
    }
\usepackage{cleveref}

\newcommand{\grillix}{\texttt{GRILLIX}}

\newcommand{\refone}[1]{#1}
\newcommand{\reftwo}[1]{#1}
\newcommand{\reftri}[1]{#1}

\makeatletter
\def\@email#1#2{%
 \endgroup
 \patchcmd{\titleblock@produce}
  {\frontmatter@RRAPformat}
  {\frontmatter@RRAPformat{\produce@RRAP{*#1\href{mailto:#2}{#2}}}\frontmatter@RRAPformat}
  {}{}
}%
\makeatother
\begin{document}

\preprint{AIP/123-QED}

\title{The impact of plasma turbulence on atomic reaction rates in \reftri{the detached ASDEX Upgrade divertor}}
\author{Konrad Eder}%
\email{konrad.eder@ipp.mpg.de}
\author{Wladimir Zholobenko}
\author{Andreas Stegmeir}%
\author{Kaiyu Zhang}
\author{Frank Jenko}
\affiliation{ 
Max Planck Institute for Plasma Physics, Boltzmannstr. 2, 85748 Garching, Germany
}%

\date{\today}

\begin{abstract}
Numerical modeling of the edge and scrape-off layer (SOL) must account for atomic processes such as hydrogenic ionization and recombination, charge-exchange, and line radiation. 
Their reaction rates depend non-linearly on density and temperature and are thus sensitive to turbulent fluctuations, whose inclusion/omission may significantly affect model outcomes.
We quantify the impact of fluctuations by studying global turbulence simulations of the edge and SOL of ASDEX Upgrade in both \reftwo{attached and detached divertor} conditions.
While the effect of fluctuations is minimal for the attached state, pronounced \reftwo{localized} discrepancies emerge in colder, detached conditions. 
\reftwo{The inclusion of turbulent fluctuations, when compared to mean-field calculations, causes a factor of 2 reduction in ionization and radiation rates local to the detachment front in the confined edge region.}
The effect arises from fluctuations crossing \reftwo{below the ionization energy threshold}, facilitated by low mean temperature and increased fluctuation amplitudes at the detachment front. 
The rate reduction (rather than rate increase) is explained by the character of divertor fluctuations (negative density-temperature correlation, i.e.~cold and dense blobs), notably distinct from characteristic fluctuations found at the outboard-midplane (positive correlation, i.e.~hot and dense blobs).
Furthermore, the cold and dense fluctuations enable efficient plasma recombination even at average temperatures above the recombination threshold.
In detached conditions, the combined plasma particle source from ionization and recombination is therefore effectively reduced by at least 50\% when compared to the standard mean-field source. 
\end{abstract}

\maketitle
\section{Introduction}
\label{sec:introduction}

In the edge and scrape off layer (SOL), plasma experiences a vast interplay of \reftwo{atomic interactions with partially un-ionized fuel and impurities}, such as ionization, recombination, charge-exchange, collisional excitation and radiative de-excitation.
These atomic reactions can act as sources/sinks of particles and energy, which have wide-reaching direct effects on plasma profiles and target fluxes \cite{stangeby2000, krasheninnikov2017, pshenov2017}.
The dependence of these atomic rates is often nonlinear with respect to plasma density and temperature, which implies that intermittent turbulent fluctuations could significantly affect the resulting rate.

This raises an issue. Much invaluable modeling and analysis of reactor scenarios is performed by transport codes, such as SOLPS-ITER\cite{wiesen2015}, EMC3 \cite{feng2014}, SOLEDGE2D \cite{bufferand2011}, and UEDGE \cite{rognlien1992}.
\refone{By design, for computational efficiency, these codes only solve for the mean background plasma quantities ($\langle n \rangle_t, \langle T \rangle_t$ etc.), usually approximating the turbulent radial transport with ad-hoc user-defined diffusivities}. 
The lack of fluctuation-related biases in the applied reaction rates may therefore introduce systematic errors in computed plasma profiles and particle/energy balance.
\reftri{Such turbulence-related rate bias is a well-known problem and has been previously investigated in analytical calculations based on SOL blob measurements \cite{krasheninnikov2009} and simulations of linear devices \cite{leddy2017}, slab-like geometries \cite{umansky2024, thrysoe2016,marandet2011, guzman2015}, and limited circular geometries \cite{fan2019}.}
\reftri{The impact of turbulence was shown to be small under attached conditions \cite{fan2019}, though it becomes significant towards detachment (at low temperature), where reaction rates are more sensitive to variations.}

Overall, the knowledge base remains limited due to a lack of comprehensive, global simulations in realistic diverted X-point geometry.
Additionally, given the fact that detached conditions are likely favored for reactor operation \cite{zohm2013, pitts2019} (where we expect more significant impacts of fluctuation on reaction rates) it is all the more relevant that further studies on the matter be conducted.
To this end, we present herein an analysis of atomic reaction rates in turbulence simulations of the edge and SOL of ASDEX Upgrade in detached X-point radiating conditions\cite{eder2025b}, performed with the full-$f$ drift-fluid code \reftwo{GRILLIX \cite{stegmeir2019,phoenix-public:grillix26}}. 

The remainder of this paper is structured as follows.
Our methodology is introduced in Section \ref{sec:demo} via a simplified toy example.
The relevant atomic reactions, as implemented in the \grillix\, code, are discussed in Section \ref{sec:reactions}, and the simulations are described in Section \ref{sec:simoverview}. 
Section \ref{sec:comparison} contains the main comparison of mean-field and turbulent rate calculations.
Our findings are then discussed in Section \ref{sec:discussion}, and placed into context with previous studies on the matter. 
Finally, we conclude our findings in Section \ref{sec:conclusion}.


\section{A demonstration of fluctuation-induced rate bias}
\label{sec:demo}

For simplicity, let us introduce the fluctuation-induced reaction rate bias by only considering the continuity equation
\begin{equation}
    \partial_t n + \nabla \cdot \bm{\Gamma} = S \,.
\end{equation}
Here, $n$ is the plasma density, $\bm{\Gamma}$ is the particle flux, and $S$ is the density source distribution. 
The latter captures reactions such as ionization and recombination, for example, which depend on local densities and temperatures.

In mean-field simulations, one is interested in obtaining the mean density profile, i.e.
\begin{equation}
    \left< \partial_t n \right>_t = 0 = \left< S - \nabla \cdot \bm{\Gamma} \right>_t \hspace{0.5cm} \Rightarrow  \hspace{0.5cm} \nabla \cdot \left< \bm{\Gamma} \right>_t = \left< S \right>_t.
\end{equation}
This is usually done by approximating the radial particle flux as (anomalous) diffusion, $\left<\Gamma_r\right>_t \approx - D_\mathrm{anom} \partial_r \left< n \right>_t$, while the real transport is due to turbulence. 
\refone{The key point to note is that on the right-hand side, there appears the average particle source $\langle S \rangle_t$, which is different from the source evaluated at average density and temperature, $S\left(\langle n \rangle_t, \langle T \rangle_t \right)$\cite{krasheninnikov2009,marandet2011,umansky2024}, as is demonstrated below.} 
However, in mean-field modeling, \reftri{the only straightforward option is} to approximate this source as the one evaluated from mean quantities, potentially introducing an averaging error.

\refone{It is illustrative to consider a minimal demonstration case at this point, following the argumentation of earlier publications \cite{krasheninnikov2009,marandet2011,umansky2024}}. 
Assume we observe the electron-impact ionization of deuterium plasma with monoatomic neutral gas, for three distinct time points, $t_1, t_2, t_3$.
The ionization source rate for electron density $n_\mathrm{e}$ roughly scales as
\begin{equation}
    S_\mathrm{iz} \propto N n_\mathrm{e} \exp{\left(- \frac{13.6\,\mathrm{eV}}{T_\mathrm{e}}\right)} \, ,
\end{equation}
depending on neutrals density $N$, electron density $n_\mathrm{e}$, and electron temperature $T_\mathrm{e}$ (expressed in eV).
Notice that $S_\mathrm{iz}$ is a strongly non-linear function \reftri{with respect to the local plasma state}, not only due to the exponential dependence on $T_\mathrm{e}$, but also due to \reftri{the presence of additional (individually linear) density dependencies}. 
If the input quantities contain fluctuations, $u(t) = \langle u \rangle_t + \tilde{u}(t), \ u \in \{ N, n_\mathrm{e}, T_\mathrm{e} \}$, the fluctuating components $\tilde{u}$ may interfere such that \reftwo{$\langle  S_\mathrm{iz}\left( N, n_\mathrm{e}, T_\mathrm{e} \right) \rangle_t \neq S_\mathrm{iz} \left( \langle N \rangle_t, \langle  n_\mathrm{e} \rangle_t, \langle  T_\mathrm{e} \rangle_t \right)$}. 

\begin{table}[htb]
\centering
\renewcommand{\arraystretch}{1.2}
\begin{tabular}{c | c c | c c}
\multicolumn{3}{c}{} & \textbf{Case 1} & \textbf{Case 2} \\
$t$ &
$N\,[10^{19}\,\mathrm{m^{-3}}]$ &
$n_\mathrm{e}\,[10^{19}\,\mathrm{m^{-3}}]$ &
$T_\mathrm{e}^{(+)}\,[\mathrm{eV}]$ & 
$T_\mathrm{e}^{(-)}\,[\mathrm{eV}]$ \\ \hline
$t_1$ & 0.1 & 0.5 & 1  & 19 \\
$t_2$ & 0.1 & 5   & 10 & 10 \\
$t_3$ & 0.1 & 9.5 & 19 & 1 \\
Mean  & 0.1 & 5   & 10 & 10 \\ \hline\hline
\multicolumn{3}{r|}{$S_\mathrm{iz} \left( \langle N \rangle_t, \langle  n_\mathrm{e} \rangle_t, \langle  T_\mathrm{e} \rangle_t \right) \,\mathrm{[ 10^{23} m^{-3}s^{-1}]}$}  &  
$5.2$ &
$5.2$ \\
\multicolumn{3}{r|}{$\langle S_\mathrm{iz} \left( N, n_\mathrm{e}, T_\mathrm{e} \right) \rangle_t \,\mathrm{[ 10^{23} m^{-3}s^{-1}]}$} &
$\bf{10.0}$ &
$\bf{2.1}$
\end{tabular}
\caption{Demonstration of the averaging bias for $S_\mathrm{iz}$. At times $t_1, t_2, t_3$, we observe neutrals density $N$, electron density $n_\mathrm{e}$, and electron temperature $T_\mathrm{e}$. Values of $T_\mathrm{e}$ are positively correlated with $n_\mathrm{e}$ in Case 1, and negatively correlated in Case 2. The final ionization source rates are computed with mean-field inputs, then with fluctuating inputs, for both cases.}
\label{tab:example_inputs}
\end{table}

To demonstrate this potential discrepancy, we evaluate $S_\mathrm{iz}$ on a synthetic fluctuating state depicted in Table \ref{tab:example_inputs}.
For simplicity, we take a constant neutrals density in this demonstration, but assume plasma density $n_\mathrm{e}$ and electron temperature $T_\mathrm{e}$ fluctuate in time $t$ as described.
We then consider two cases with different temperature fluctuations (but identical densities $N, n_\mathrm{e}$).
In the first, temperature rises simultaneously with electron density. In other words, they are perfectly positively correlated with a Pearson correlation coefficient of $R_{n_\mathrm{e}, T_\mathrm{e}} = +1$.  
In the second case, temperature is anti-correlated instead ($R_{n_\mathrm{e}, T_\mathrm{e}} = -1$), though note that the time average $\langle T_\mathrm{e} \rangle_t$ remains identical.

The final two rows of Table \ref{tab:example_inputs} depict the resultant ionization sources for both cases, computed from mean quantities $\langle u \rangle_t$, and the fluctuating quantities $u(t)$ (and averaged afterwards).
Even in such a simple case, two effects are immediately evident.
The ionization rate varies by a factor $\gtrsim 2$ depending on the averaging order, and the direction of change (whether the fluctuation-including rate increases or decreases) depends on the correlation between density and temperature.

While this example is deliberately constructed to emphasize the effect of fluctuations, it is nonetheless representative of realistic conditions, since fluctuations of order unity are commonly observed in the SOL \cite{ritz1989,zweben2007,garcia2007,boedo2009}.
As we will discuss in the following sections, similar discrepancies emerge in our global turbulence simulations of ASDEX Upgrade in detached X-point radiator conditions \cite{eder2025b}, which were validated against the experiment and thus appear to be quite realistic.

\section{Reactions in the \grillix\, code}
\label{sec:reactions}

The turbulence simulations discussed hereafter have been performed with the \grillix\, code \cite{stegmeir2019,phoenix-public:grillix26}, which consists of a \mbox{full-$f$} drift-fluid plasma model \cite{zholobenko2024} coupled to a mono-atomic neutral gas model \cite{eder2025a} based on the advanced fluid neutrals (AFN) approach \cite{horsten2017, uytven2020}. \reftwo{An overview of the full model is given in the Appendix.} 
\begin{table}[htb]
\centering
\renewcommand{\arraystretch}{1.2}
\begin{tabular}{l r}
\hline
    Electron impact ionization & $\mathrm{e}^- + \mathrm{D}^0 \longrightarrow 2 \mathrm{e}^- + \mathrm{D}^+$ \\ \hline 
    3-body recombination & $2 \mathrm{e}^{-} + \mathrm{D}^{+} \longrightarrow \mathrm{e}^{-} + \mathrm{D}^{0}$ \\
    Radiative recombination & $\mathrm{e}^{-} + \mathrm{D}^{+} \longrightarrow \mathrm{D}^{0} + \gamma$ \\ \hline 
    Charge exchange & $\mathrm{D}^{+} + \mathrm{D}^{0} \longrightarrow \mathrm{D}^{0} + \mathrm{D}^{+}$ \\
\hline
\end{tabular}
\caption{Atomic processes modeled in the \grillix\, turbulence simulations.}
\label{tab:reactions}
\end{table}

The two model components interact through the reactions outlined in Table \ref{tab:reactions}.
Ionization ('iz') and (3-body + radiative) recombination ('rc') result in non-linear sources and sinks of plasma density, defined as
\begin{align}
    S_\mathrm{iz} &= N n_\mathrm{e} \reftri{k_\mathrm{iz}(n_\mathrm{e}, T_\mathrm{e})} \, , \label{eq:siz}\\
    S_\mathrm{rc} &= - n_\mathrm{e} n_\mathrm{i} \reftri{k_\mathrm{rc}(n_\mathrm{e}, T_\mathrm{e})} \, , \label{eq:src}
\end{align}
with neutrals density $N$, electron density $n_\mathrm{e}$, deuterium ion density $n_\mathrm{i}$, and the corresponding reaction rate coefficients $k = \langle \sigma v_\mathrm{rel} \rangle$ taken from the AMJUEL database \cite{julichdata}.
\reftri{Notice that AMJUEL coefficients $k_\mathrm{iz}, k_\mathrm{rc}$ contain minor density dependencies, $k_\mathrm{iz} \propto n_\mathrm{e}^{0.12}$ and $k_\mathrm{rc} \propto n_\mathrm{e}^{0.02}$ in the range of $n_\mathrm{e} \in [10^{18}, 10^{21}]\,\mathrm{m^{-3}}$}.
\reftwo{We assume quasi-neutrality of the plasma, thus $n_\mathrm{e} = n_\mathrm{i} = n$}.

Additionally, \grillix\, models impurity radiation in the coronal approximation \cite{stroth2022xpr,eder2025b} with fixed-fraction impurity content (i.e.~impurity density $n_\mathrm{imp}$ is a fixed fraction of local ion density, $n_\mathrm{imp} = c_\mathrm{imp} n_\mathrm{i}$). The total radiation density reads
\begin{equation}\label{eq:prad}
    P_\mathrm{rad} = n_\mathrm{imp} n_\mathrm{e} L_Z\left( T_\mathrm{e} \right) \, ,
\end{equation}
with the radiation rate coefficient $L_Z$ containing line radiation and bremsstrahlung. 

\reftwo{At simulation runtime, the ionization / recombination / radiation rates $S (n, T_\mathrm{e}, N)  \in \{ S_\mathrm{iz}, -S_\mathrm{rc}, P_\mathrm{rad} \}$ in equations \eqref{eq:siz}, \eqref{eq:src}, \eqref{eq:prad} are updated at every simulation time step, i.e.~no sub-cycling is applied.}

\section{Simulation overview}
\label{sec:simoverview}

\begin{figure}[htbp]
    \centering
    \includegraphics[width=1.0\linewidth,trim={0 20 30 10},clip=true]{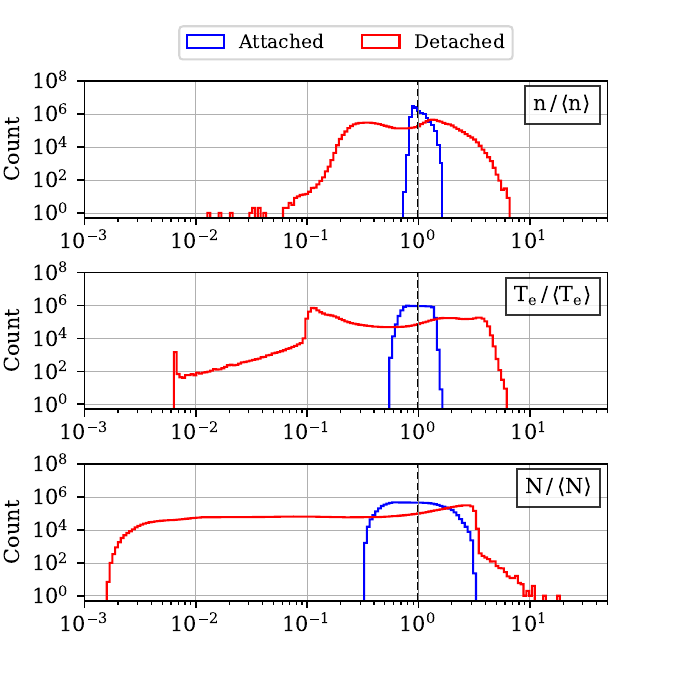}
    \caption{Histograms of normalized plasma density, electron temperature, and neutrals density values in the control volume $V_\mathrm{xpt}$ for both simulation cases. \reftri{Samples are taken from 100 equidistant intervals in log-space from $10^{-3}$ to $10^{2}$.}}
    \label{fig:1d_hist}
\end{figure}
We will consider two simulations of ASDEX Upgrade (AUG), which we characterize as attached and detached cases, respectively.
The former is based on AUG L-mode discharge \#38839 at $1.3\,\mathrm{s}$, performed in favorable configuration at toroidal field $B_\mathrm{tor} = - 2.5\,\mathrm{T}$ and plasma current $I_p = 0.8\,\mathrm{MA}$. 
GRILLIX was previously analyzed and validated against the discharge \cite{eder2025a}, and our simulation parameters are chosen specifically matching the '3-moment, Dirichlet boundary' case documented in Section 5.1.~of Ref.~\cite{eder2025a}.
The simulation was performed without impurity radiation $(P_\mathrm{rad} = 0)$ and features modest fluctuation amplitudes as outlined in the histogram of normalized $n, T_\mathrm{e}, N$ values shown in Figure \ref{fig:1d_hist}.
Note that the histogram samples are obtained from a particular control volume $V_\mathrm{xpt}$ above the X-point. This choice will be discussed later on, see Section \ref{sec:comparison}.

The second, detached simulation, is based on AUG discharge \#40333 at $2.4\,\mathrm{s}$ featuring an X-point radiator \cite{bernert2025} (in L-mode conditions), with toroidal field $B_\mathrm{tor} = -2.4\,\mathrm{T}$ and plasma current $I_p = 0.8\,\mathrm{MA}$.
\reftwo{Only in this case, nitrogen impurity radiation is enabled, whereby we assume an impurity concentration $c_\mathrm{imp} = 5\%$.
This choice was motivated by previous extensive validation against the discharge \cite{eder2025b}, and our simulation parameters match those of the ``High-XPR'' case described in Section 3.1.~of Ref.~\cite{eder2025b}.}
In contrast to the attached simulation, this case features a detachment front raised above the X-point, and significantly amplified fluctuations of density and temperature, as shown in Figure \ref{fig:1d_hist}. 
Here, we observe that individual fluctuations near the X-point reach amplitudes of up to \reftwo{$500\%$ of their mean value}. 

For both simulations, the computational domain spans 16 poloidal planes, with each plane comprising Cartesian $(R,Z)$ grids with a resolution of $(1.44\,\mathrm{mm} \times 1.44\,\mathrm{mm})$ per grid point.
The simulation domains are limited in radial direction by normalized flux surface labels $\rho_\mathrm{pol}$, $\left[ 0.90, 1.035 \right]$ for the attached case and $\left[ 0.90, 1.04 \right]$ for the detached case.
\reftwo{Attached and detached simulations are run for $7.7\,\mathrm{ms}$ and $3.5\,\mathrm{ms}$, where we consider the 50 last snapshots of each simulation (spanning $0.12\,\mathrm{ms}$) for subsequent analysis.
Turbulent fluctuations in both simulations are measured to have a decorrelation time \mbox{$t_c\lesssim 15\,\mathrm{\mu s}$}, evaluated at $\rho_\mathrm{pol}=0.98$ near the X-point.\\
We then obtain a `fluctuation-including' rate-average, denoted $\langle S \rangle$, by averaging the sources $S$ over toroidal angle (spanning all 16 poloidal planes) and the time window specified above.
To produce comparable 'mean-field' rates, we average first the input fields $(n, T_\mathrm{e}, N)$, resulting in a smoothed plasma state without intermittent turbulent structures.
These smoothed quantities are then used to re-evaluate equations \eqref{eq:siz}, \eqref{eq:src}, \eqref{eq:prad}, with the resulting source rates denoted as $S \langle \circ \rangle$,.}

We remark that the mean-field source $S \langle \circ \rangle$ obtained in this manner a posteriori may not reflect a realistic source for the given simulation state; if a full simulation were conducted using the mean-field rates $S \langle \circ \rangle$ instead of fluctuation-including $\langle S \rangle$, a different particle balance may emerge.
Indeed, conducting a "perfect" comparison of turbulent and mean field rates is challenging in that sense.
In principle, a consistent comparison could be achieved by a coupled iteration scheme of turbulence and mean-field codes. The turbulent plasma state computed by the former may be averaged and then fed to a transport solver, which then determines a new particle balance on transport timescales. Such couplings have been successfully demonstrated both for the core and edge regions \cite{disiena2022, zhang2019}, though we consider such an exercise beyond the scope of this paper. 

\section{Comparison of turbulent and mean-field rates}
\label{sec:comparison}

\begin{figure*}[htb]
    \centering
    \includegraphics[width=1\linewidth]{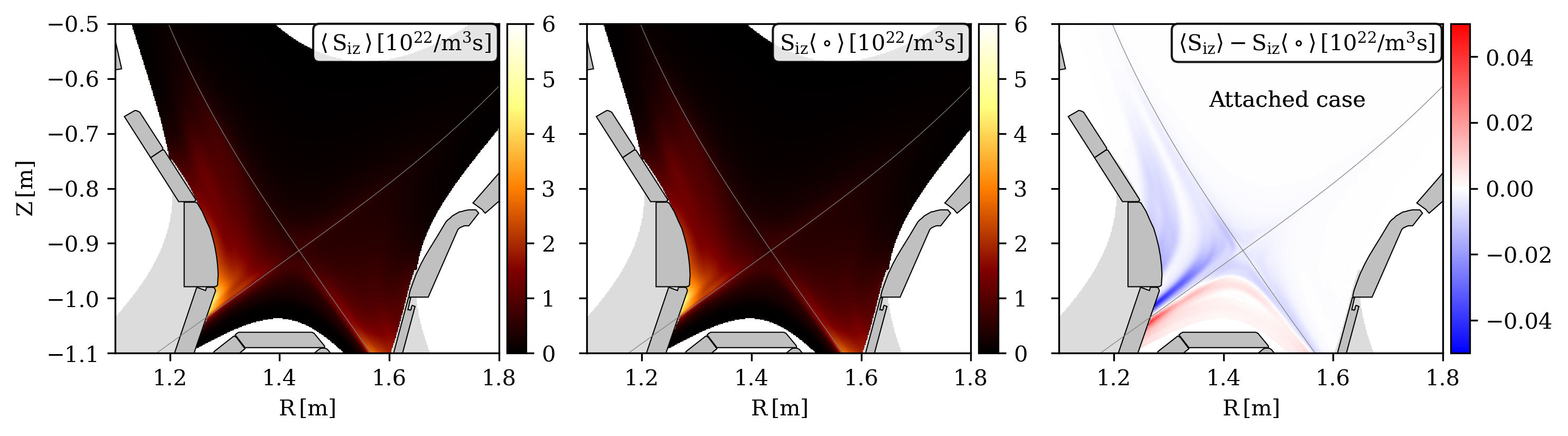}
    \includegraphics[width=1\linewidth]{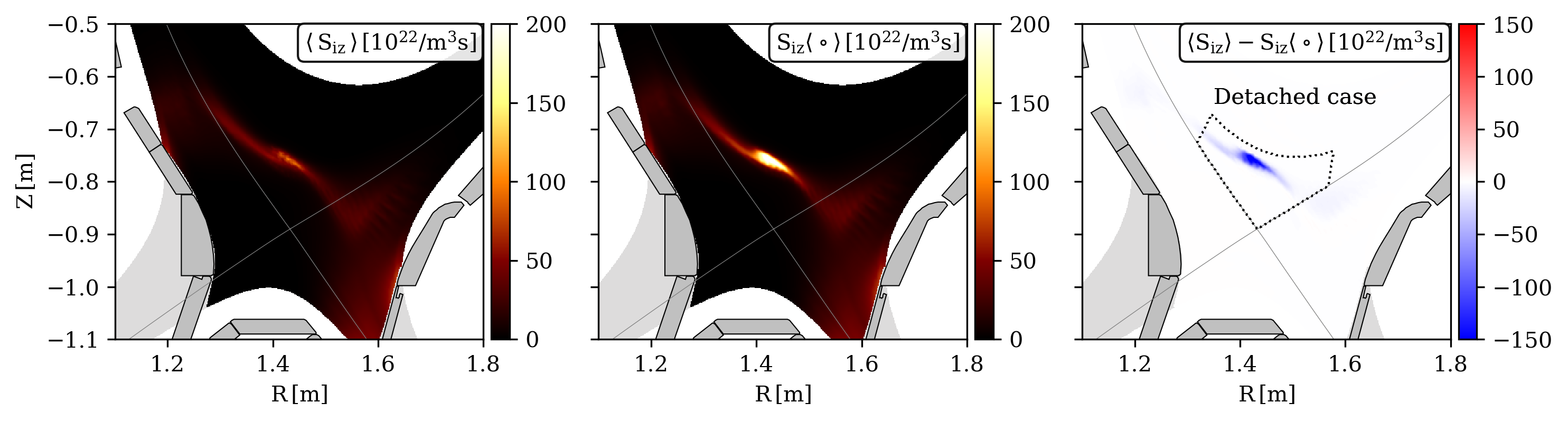}
    \caption{Ionization source in the attached (top row) and detached (bottom row) simulations, averaged over time and toroidal angle. Left column: the rate is first computed from fluctuating input fields and then averaged. Middle column: the rate is computed on pre-averaged (smooth) fields. Right column: absolute difference between the two evaluations.}
    \label{fig:s_iz_2d_detached}
\end{figure*}

We begin with the averaged ionization rates $\langle S_\mathrm{iz} \rangle$ and $S_\mathrm{iz} \langle \circ \rangle$ for the attached case, shown in Figure \ref{fig:s_iz_2d_detached}, top row. 
Ionization occurs close to the targets, and at very similar rates regardless of averaging order.
The absolute difference $\langle S_\mathrm{iz} \rangle - S_\mathrm{iz} \langle \circ \rangle$ equals $4\cdot 10^{19}\,\mathrm{m^{-3}s^{-1}}$ at most, resulting in a relative difference of approximately 1\%. 
A more pronounced difference emerges in the detached case, shown in the bottom row of Figure \ref{fig:s_iz_2d_detached}.
In this case, the ionization region is shifted upward into the confined region, forming a densely ionizing spot above the X-point (corresponding to the location of the X-point radiator).
There, fluctuation-including and mean-field rates respectively peak at $\sim 1\cdot 10^{23}$ and $2.5\cdot 10^{23}\,\mathrm{m^{-3}s^{-1}}$ leading to a factor 2.5 amplification relative to the fluctuation-including rate.

\begin{figure*}[htb]
    \centering
    \includegraphics[width=1\linewidth]{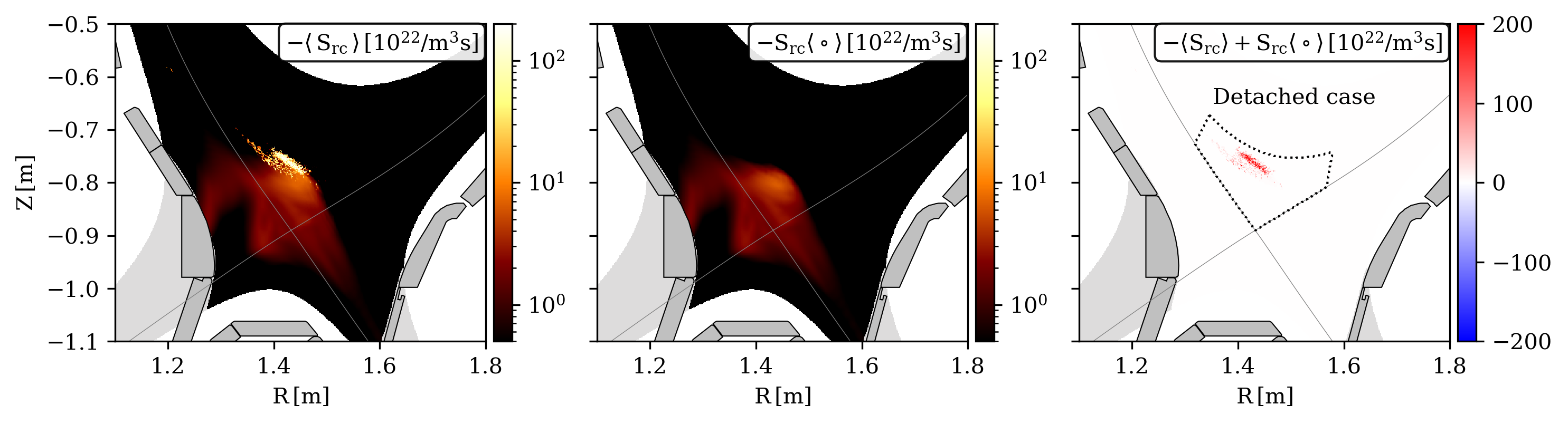}
    \includegraphics[width=1\linewidth]{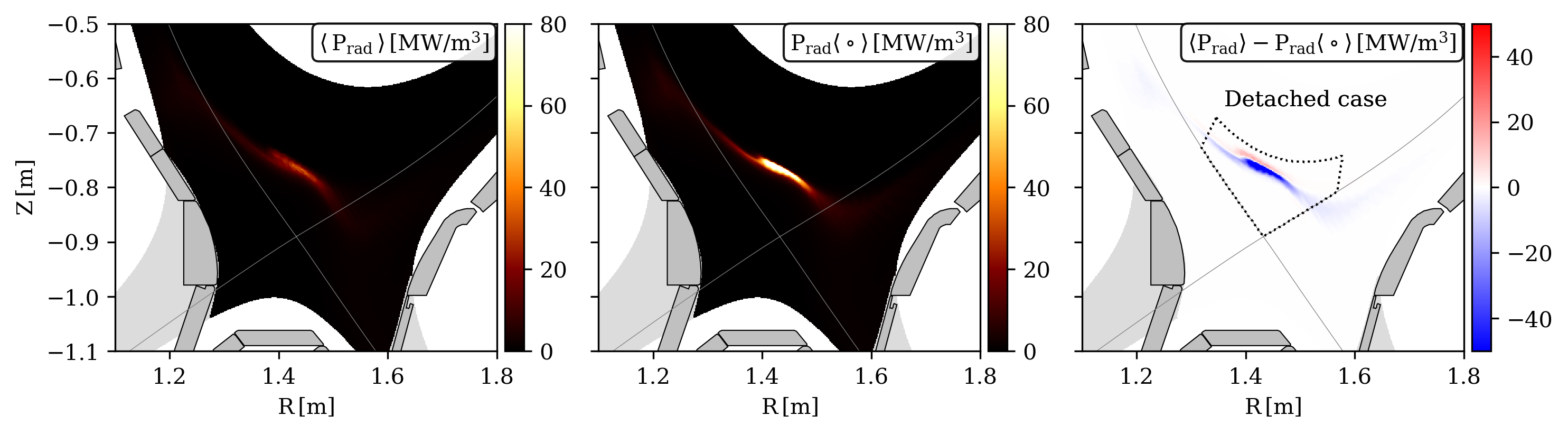}
    \caption{Recombination sink (top row) and impurity radiation density (bottom row), both in the detached simulation, averaged over time and toroidal angle. Left column: the rate is first computed from fluctuating input fields and then averaged. Middle column: the rate is computed on pre-averaged (smooth) fields. Right column: absolute difference between the two evaluations. \reftri{Note the logarithmic colorbar for showing recombination sink rates.}}
    \label{fig:s_rc_p_rad_2d}
\end{figure*}

We perform an analogous comparison of recombination sink rates $S_\mathrm{rc}$, shown in the top row of Figure \ref{fig:s_rc_p_rad_2d}. 
We show only the detached case here, as divertor temperatures in the attached reference case are too high for plasma to recombine ($S_\mathrm{rc} \sim 0$ regardless of averaging order). 
The fluctuation-including recombination source $\langle S_\mathrm{rc} \rangle $ is strongly localized (corresponding to the center of the X-point radiator structure), with individual fluctuations (locally cold and dense structures) facilitating local bursts of recombination \cite{eder2025b}. 
The mean-field evaluation $S_\mathrm{rc} \langle \circ \rangle$ on the other hand yields very low rates, smaller than the fluctuation-including rate by two orders of magnitude.

Lastly, the bottom row of Figure \ref{fig:s_rc_p_rad_2d} compares the effective impurity radiation rates.
We again show only the detached case, as the attached simulation was performed without the impurity model, thus $P_\mathrm{rad} = 0$.
The resulting differences closely match the comparison of ionization rates, as we find that the mean-field evaluation $P_\mathrm{rad} \langle \circ \rangle$ yields systematically higher rates compared to the fluctuation-including $\langle P_\mathrm{rad} \rangle$. 
At the location corresponding to maximal $\langle P_\mathrm{rad} \rangle$, the mean-field variant $P_\mathrm{rad} \langle \circ \rangle$ exceeds it by more than a factor of 3.

We find that differences between fluctuation-including and mean-field rates become significant only in the detached simulation, \reftwo{and are highly localized to the detachment front within the confined plasma edge region therein}.
\reftri{To quantify the observed discrepancy concisely}, we propose two averaging procedures, taken over a control volume  $V_\mathrm{xpt}$ outlining the region where $\langle \reftri{S} \rangle - \reftri{S} \langle \circ \rangle$ takes significant non-zero values in the detached case.
$V_\mathrm{xpt}$ is thus defined as a toroidally continuous volume above the X-point, bounded by poloidal angles $\theta_\mathrm{pol} \in [-2.042\,\mathrm{rad}, -1.728\,\mathrm{rad}]$ and normalized flux surfaces $\rho_\mathrm{pol} \in [0.98, 1.0]$.
The margins of $V_\mathrm{xpt}$ are shown in dotted lines in the rightmost subplots of figures \ref{fig:s_iz_2d_detached} and \ref{fig:s_rc_p_rad_2d}. 

First, we consider a simple supremum norm of rates $\reftri{S}$, where we evaluate the ratio of maximal turbulent and mean-field rates in $V_\mathrm{xpt}$,
\begin{equation}\label{eq:norm_sup}
    \frac{\left \Vert \langle \reftri{S} \rangle \right \Vert_\mathrm{sup} }{\left \Vert \reftri{S} \langle \circ \rangle\right \Vert_\mathrm{sup} } = \frac{\max \left( \langle \reftri{S} \rangle \right)  \big\vert_{V_\mathrm{xpt}}}{\max \left( \reftri{S} \langle \circ \rangle\right) \big\vert_{V_\mathrm{xpt}}}, \quad \reftri{S}  \in \{ S_\mathrm{iz}, -S_\mathrm{rc}, P_\mathrm{rad} \} \, ,
\end{equation}
and additionally, the ratio of volume averages taken over $V_\mathrm{xpt}$,
\begin{equation}\label{eq:norm_vol}
    \frac{\left \Vert \langle \reftri{S} \rangle \right \Vert_\mathrm{avg}}{\left \Vert \reftri{S} \langle \circ \rangle \right \Vert_\mathrm{avg}} = \frac{\int_{V_\mathrm{xpt}} \langle \reftri{S} \rangle_{\varphi, t} \mathrm{d} V}{\int_{V_\mathrm{xpt}} \reftri{S} \langle \circ \rangle \mathrm{d}V}, \quad \reftri{S}  \in \{ S_\mathrm{iz}, -S_\mathrm{rc}, P_\mathrm{rad} \} \, ,
\end{equation}
which is identical to the ratio of volume integrals.
Note that for all $\reftri{S}  \in \{ S_\mathrm{iz}, -S_\mathrm{rc}, P_\mathrm{rad} \}$ it holds that $\reftri{S} \geq 0$, thus we are not required to take absolute values.

The resulting norms for each reaction rate are recorded in Table \ref{tab:ratios_xpr}. 
Regarding the attached case, we find fluctuation-including and mean-field rates to be consistently similar, with ratios $\sim 1$ for all reactions considered.
Let us recall that no impurities were modeled in the attached simulation ($P_\mathrm{rad} = 0$), and thus no ratios are provided there.
In the detached case, on the other hand, we identify that turbulence averaged rates reduce significantly, locally by up to $\sim 60\%$ for ionization and $\sim 70\%$ for impurity radiation. 
Recombination in the turbulence simulations is concentrated on only a small set of points, whereas it covers a broader region in the mean-field approach (at a much reduced rate). Consequently, despite the extreme supremum ratio ($>50$), the relative rate difference over the $V_\mathrm{xpt}$ volume average reduces to only a factor of $\sim 4$.

\begin{table}[htb]
    \centering
    \begin{tabular}{l|ll|ll}
        ~ & \multicolumn{2}{c|}{\textbf{Attached case}} & \multicolumn{2}{c}{\textbf{Detached case}} \\ 
        ~ & $V_\mathrm{xpt}$ sup & $V_\mathrm{xpt}$ avg & $V_\mathrm{xpt}$ sup & $V_\mathrm{xpt}$ avg \\ \hline
        $\langle S_\mathrm{iz} \rangle \, / \, S_\mathrm{iz}\langle \circ \rangle$ & 0.995 & 0.996 & 0.367 & 0.653 \\
        $\langle S_\mathrm{rc} \rangle \, / \, S_\mathrm{rc}\langle \circ \rangle$ & 1.03 & 1.01 & 59.2 & 4.40 \\
        $\langle P_\mathrm{rad} \rangle \, / \, P_\mathrm{rad}\langle \circ \rangle$ & - & - & 0.274 & 0.602 \\\hline
        ~ & \multicolumn{2}{c|}{$R_{n, T_\mathrm{e}} = -0.73$} & \multicolumn{2}{c}{$R_{n, T_\mathrm{e}} = -0.46$} 
    \end{tabular}
    \caption{Ratios of reaction rates obtained from fluctuating inputs, $\langle \reftri{S} \rangle$, relative to rates obtained from mean field inputs, $\reftri{S} \langle \circ \rangle$. Values smaller than 1 indicate that including fluctuations reduces the effective reaction rate, and vice versa. For both simulation cases, we compute the ratios in the control volume $V_\mathrm{xpt}$ from equations \eqref{eq:norm_sup} ('sup') and \eqref{eq:norm_vol} ('avg'), respectively. The attached simulation was performed without impurity radiation $P_\mathrm{rad}$. We additionally record the Pearson correlation coefficient $R_{n, T_\mathrm{e}}$ of density and electron temperature samples in $V_\mathrm{xpt}$.}
    \label{tab:ratios_xpr}
\end{table}

\section{Discussion}
\label{sec:discussion}

\begin{figure*}[htb]
    \centering
    \includegraphics[width=\linewidth,trim={0 0 0 0},clip]{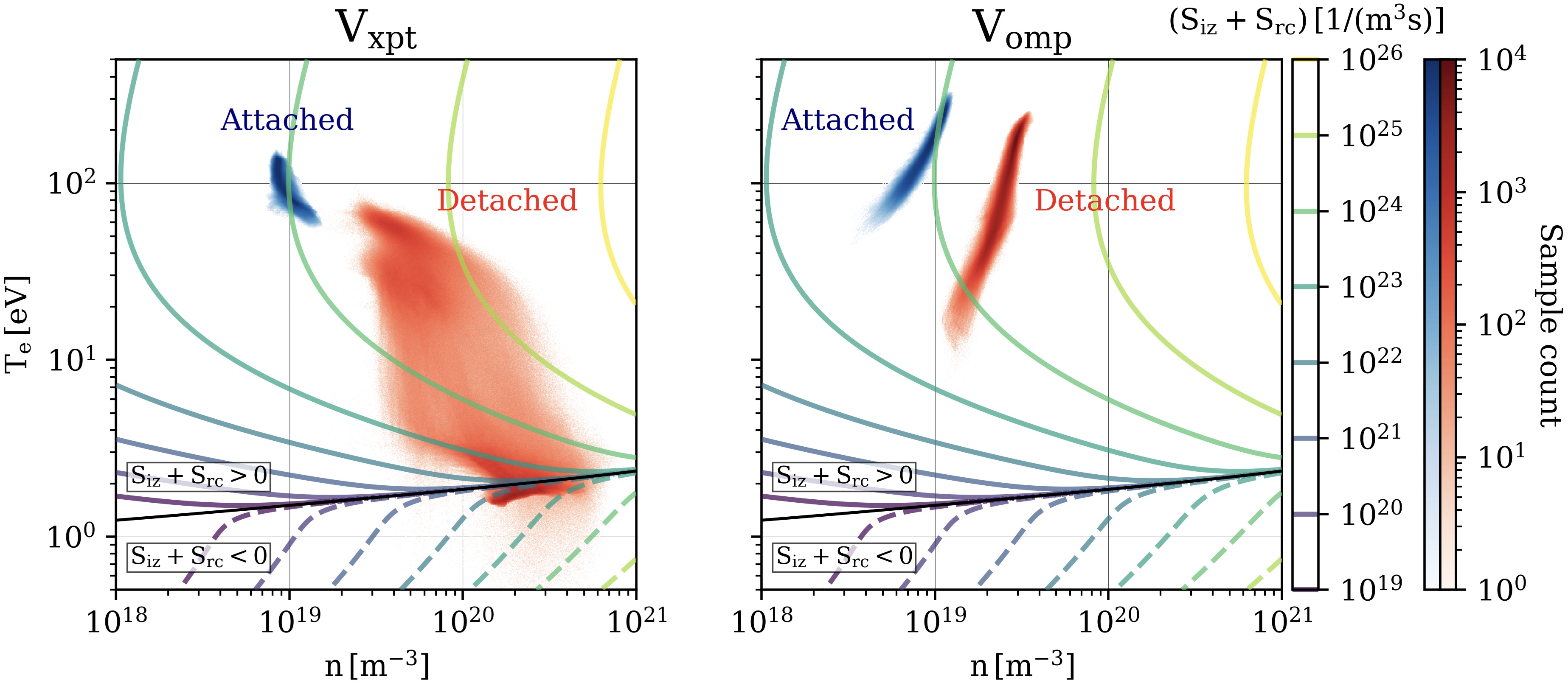}
    \caption{\reftri{2D-histograms of $(n, T_\mathrm{e})$ samples in the attached (blue) and detached (red) simulations, taken from $800 \times800 $ equidistant intervals in log-space between the shown axis boundaries. Overlaid are contours of the total density source rate $S_\mathrm{iz} + S_\mathrm{rc}$, assuming a neutrals density of $N = 2.5 \cdot 10^{18}\,\mathrm{m^{-3}}$. Solid contours indicate positive values, $S_\mathrm{iz} + S_\mathrm{rc} > 0$, where the ionization dominates. Dashed contours indicate negative values, $S_\mathrm{iz} + S_\mathrm{rc} < 0$, where recombination dominates.} The left subplot indicates samples taken from the X-point control volume $V_\mathrm{xpt}$, the right subplot shows samples from the OMP control volume $V_\mathrm{omp}$.}
    \label{fig:2d_hist_with_contours}
\end{figure*}

To get a better understanding of how plasma fluctuations may shape the effective reaction rates, let us switch to a 2-dimensional view of the distributions of $(n, T_\mathrm{e})$ samples in $V_\mathrm{xpt}$, shown in the left subplot of Figure \ref{fig:2d_hist_with_contours}.
\reftri{We further overlay contour lines of the total density source $S_\mathrm{iz} + S_\mathrm{rc}$, evaluated at the respective location in $(n, T_\mathrm{e})$ space. 
Since the ionization source \eqref{eq:siz} also depends linearly on neutrals density (a dimension not captured in the Figure), the contour overlay assumes $N = 2.5 \cdot 10^{18}\,\mathrm{m^{-3}}$.
This corresponds to the average neutrals density of samples with electron temperature $T_\mathrm{e} < 5\,\mathrm{eV}$.}

\reftri{Note how the contour density increases for low temperatures $T_\mathrm{e} \sim 3\,\mathrm{eV}$, highlighting electron temperature as the dominant dependency.
For temperatures $T_\mathrm{e} > 5\,\mathrm{eV}$, the contours are determined only by $S_\mathrm{iz}$, since $S_\mathrm{rc} \sim 0$.
At $T_\mathrm{e} \sim 2\,\mathrm{eV}$, the recombination sink becomes significant (while the ionization source diminishes), resulting in $S_\mathrm{iz} + S_\mathrm{rc} < 0$.
The threshold contour where $S_\mathrm{iz} + S_\mathrm{rc} = 0$ is highlighted in black.}
 
Fluctuations in the attached case are situated in a relatively narrow location in $(n, T_\mathrm{e})$ phase-space, resulting in a similarly narrow range of $S_\mathrm{iz}$ values ($ 10^{24} - 10^{25}\,\mathrm{m^{-3}s^{-1}}$).
Regardless of whether one considers $\langle S_\mathrm{iz} \rangle$ or $S_\mathrm{iz} \langle \circ \rangle$, we can intuit that both orderings will yield similar rates of $\sim 5\cdot 10^{24}\,\mathrm{m^{-3}s^{-1}}$, since $S_\mathrm{iz}$ changes little in the range of interest and can thus be treated as \textit{almost} linear.

Fluctuations in the detached case are clustered at lower temperatures and span a broader phase-space area.
While the average electron temperature of the population is still above the threshold temperature, $\langle T_\mathrm{e} \rangle \sim 15\,\mathrm{eV}$, a significant fraction of samples \reftri{cross into the threshold region $T_\mathrm{e} \lesssim 3\mathrm{eV}$ spanning multiple orders of magnitude with respect to $S_\mathrm{iz}$}.
Here, we cannot intuitively determine the final source rate and must instead rely on proper averaging over the fluctuations. 

\subsection{Fluctuations enable recombination}

The dramatic differences in recombination rate shown in Figure \ref{fig:s_rc_p_rad_2d} and Table \ref{tab:ratios_xpr} can be understood by again considering the distributions in Figure \ref{fig:2d_hist_with_contours}.
Recombination becomes significant primarily toward cold plasmas $(\lesssim 3\,\mathrm{eV}$), replacing ionization.
The mean-field evaluation $S_\mathrm{rc}\langle \circ \rangle$ takes as averaged inputs for the detached case $\langle n \rangle \sim 10^{20}\,\mathrm{m^{-3}}$ and $\langle T_\mathrm{e} \rangle \sim 15\,\mathrm{eV}$.
This is far above the required temperature for recombination, yielding $S_\mathrm{rc} \langle \circ \rangle \sim 0$.

Considering the fluctuation-including evaluation $\langle S_\mathrm{rc} \rangle$, this stringent temperature condition for recombination is fulfilled in the detached case, though only for a subset of samples (which correspond to the intermittent bursts of recombination observed at the turbulent detachment front).
Therefore, even as $S_\mathrm{rc} \sim 0$ holds for the population of fluctuations above the threshold, this results in a final non-zero average $\langle S_\mathrm{rc} \rangle > 0$.
Notice that the ratio $\langle S_\mathrm{rc} \rangle / S_\mathrm{rc} \langle \circ \rangle$ is then somewhat ill-defined, and thus the recombination ratios provided in Table \ref{tab:ratios_xpr} should be considered with care.

\subsection{Role of density-temperature correlations}

Table \ref{tab:ratios_xpr} further shows that the fluctuation-including ionization rate is \textit{reduced} compared to the mean-field rate. 
This is a somewhat surprising result, as previous analyses on the impact of turbulent fluctuations \cite{guzman2015, thrysoe2016, leddy2017, fan2019, umansky2024} find that including fluctuations results in an effective ionization \textit{increase} as opposed to a decrease.
However, Ref.~\cite{leddy2017, thrysoe2016} also explicitly point out that analysis is performed on turbulent plasma blobs, that is, plasma fluctuations where \refone{density and temperature tend to be positively correlated \cite{Kube2019,zhang2024}}. 
As we have pointed out in the demonstration case in Section \ref{sec:demo}, such correlations \reftri{(in combination with large fluctuations)} will significantly impact the final ionization rate.

We can estimate from Figure \ref{fig:2d_hist_with_contours} that $n$ and $T_\mathrm{e}$ in the X-point volume are negatively correlated, i.e., high temperatures are most associated with low densities and vice-versa.
We find this for both simulations, though as mentioned above, the effective ionization rate is only impacted if fluctuations approach the critical low-temperature region $T_\mathrm{e} \lesssim 3\,\mathrm{eV}$.
Computing the Pearson correlation coefficients $R_{n, T_\mathrm{e}}$ for the control volume $V_\mathrm{xpt}$ yields $-0.73$ and $-0.46$ for the attached and detached simulation, respectively (also shown in Table \ref{tab:ratios_xpr}).

For further comparison, let us also consider fluctuation samples taken at the outboard-midplane (OMP), shown in the right subplot of Figure \ref{fig:2d_hist_with_contours}. More precisely, we define an analogous OMP control volume $V_\mathrm{omp}$ as a toroidally continuous region bounded by poloidal angles $\theta_\mathrm{pol} \in [-0.75\,\mathrm{rad}, 0.75\,\mathrm{rad}]$ (symmetric around the midplane) and normalized flux surfaces $\rho_\mathrm{pol} \in [ 0.94, 1.02]$.
There, we find that both fluctuation populations are similar and positively correlated, with correlation coefficients $R_{n, T_\mathrm{e}} = +0.78$ and $+0.74$ for the attached and detached simulations, respectively. 
Since both OMP volume populations are far above the temperature threshold, we would expect no significant difference between $\langle S_\mathrm{iz} \rangle$ and $S_\mathrm{iz} \langle \circ \rangle$.
Indeed, performing the same calculation as in Table \ref{tab:ratios_xpr} yields ratios close to 1, which we document in Table \ref{tab:ratios_omp}.

\begin{table}[htb]
    \centering
    \begin{tabular}{l|ll|ll}
        ~ & \multicolumn{2}{c|}{\textbf{Attached case}} & \multicolumn{2}{c}{\textbf{Detached case}} \\ 
        ~ & $V_\mathrm{omp}$ sup & $V_\mathrm{omp}$ avg & $V_\mathrm{omp}$ sup & $V_\mathrm{omp}$ avg \\ \hline
        $\langle S_\mathrm{iz} \rangle \, / \, S_\mathrm{iz}\langle \circ \rangle$ & 0.997 & 0.997 & 0.983 & 0.991 \\
        $\langle S_\mathrm{rc} \rangle \, / \, S_\mathrm{rc}\langle \circ \rangle$ & 1.00 & 1.00 & 1.03 & 1.02 \\
        $\langle P_\mathrm{rad} \rangle \, / \, P_\mathrm{rad}\langle \circ \rangle$ & - & - & 0.963 & 1.05 \\\hline
        ~ & \multicolumn{2}{c|}{$R_{n, T_\mathrm{e}} = +0.78$} & \multicolumn{2}{c}{$R_{n, T_\mathrm{e}} = +0.74$} 
    \end{tabular}
    \caption{Same as Table \ref{tab:ratios_xpr}, evaluated for the control volume $V_\mathrm{omp}$ defined at the outboard-midplane.}
    \label{tab:ratios_omp}
\end{table}

However, let us recall that in our simulations, the computational domains are bounded by flux surfaces (as defined in Section \ref{sec:simoverview}), and we neglect main-chamber recycling (and thus ionization) in the far-SOL. 
These may be significant \cite{zito2021}, however, and we could thus expect plasma fluctuations to significantly influence ionization rates there as well.

\subsection{Further exploration on synthetic fluctuations}

At this point, it remains to be seen how differently correlated fluctuations (assuming they cross into the low temperature threshold) would affect the final ionization rate.
To investigate this further, we perform two numerical experiments by constructing synthetic fluctuation samples with no correlation $R_{n, T_\mathrm{e}} = 0$ and positive correlation $R_{n, T_\mathrm{e}} > 0$.
To this end, we again take the samples obtained from the X-point volume (where the detached case reaches the threshold), and shuffle $N, n, T_\mathrm{e}$ samples in time and toroidal angle (independent of each other).
The resulting population is fully decorrelated, yielding $R_{n, T_\mathrm{e}} = 0$, while preserving their averaged values $\langle N \rangle, \langle n \rangle, \langle T_\mathrm{e} \rangle$ and 1-dimensional distributions from Figure \ref{fig:1d_hist}.
Additionally, by arranging $N, n, T_\mathrm{e}$ samples in monotonically increasing order (independent of each other), we can construct a positively correlated distribution of samples.
We then repeat the rate evaluation from Table \ref{tab:ratios_xpr} for these two modified input sets and record the results in Table \ref{tab:ratios_xpr_modified}.
\begin{table}[htb]
    \centering
    \begin{tabular}{l|ll|ll}
        ~ & \multicolumn{2}{c|}{\textbf{Attached case}} & \multicolumn{2}{c}{\textbf{Detached case}} \\ 
        ~ & $V_\mathrm{xpt}$ sup & $V_\mathrm{xpt}$ avg & $V_\mathrm{xpt}$ sup & $V_\mathrm{xpt}$ avg \\ \hline
        \textbf{Decorrelated inputs} & \multicolumn{2}{c|}{~} & \multicolumn{2}{c}{~} \\ 
        $\langle S_\mathrm{iz} \rangle \, / \, S_\mathrm{iz}\langle \circ \rangle$ & 1.00 & 1.00 & 0.931 & 1.09 \\
        $\langle S_\mathrm{rc} \rangle \, / \, S_\mathrm{rc}\langle \circ \rangle$ & 1.02 & 1.01 & 150 & 5.73 \\ 
        $\langle P_\mathrm{rad} \rangle \, / \, P_\mathrm{rad}\langle \circ \rangle$ & - & - & 0.418 & 0.753 \\ \hline
        ~ & \multicolumn{2}{c|}{$R_{n, T_\mathrm{e}} = 0.0$} & \multicolumn{2}{c}{$R_{n, T_\mathrm{e}} = 0.0$} \\ \hline\hline
        \textbf{Ordered inputs} & \multicolumn{2}{c|}{~} & \multicolumn{2}{c}{~} \\ 
        $\langle S_\mathrm{iz} \rangle \, / \, S_\mathrm{iz}\langle \circ \rangle$ & 1.01 & 1.01 & 3.37 & 2.72 \\
        $\langle S_\mathrm{rc} \rangle \, / \, S_\mathrm{rc}\langle \circ \rangle$ & 1.00 & 1.00 & 2.52 & 1.11 \\ 
        $\langle P_\mathrm{rad} \rangle \, / \, P_\mathrm{rad}\langle \circ \rangle$ & - & - & 0.609 & 0.950 \\ \hline
        ~ & \multicolumn{2}{c|}{$R_{n, T_\mathrm{e}} = +0.98$} & \multicolumn{2}{c}{$R_{n, T_\mathrm{e}} = +0.93$} 
    \end{tabular}
    \caption{Same as Table \ref{tab:ratios_xpr}, but evaluated with synthetic input quantities that have been deliberately decorrelated (first row group) and positively correlated (second row group).}
    \label{tab:ratios_xpr_modified}
\end{table}
Once more, no differences are found for the attached case (for both synthetic populations), identical to the first round of evaluations with true plasma inputs.
For the detached case, the act of decorrelation raises fluctuation-including rates $\reftwo{\langle S \rangle}$ considerably, with the turbulent-to-mean-field ratio $\langle S_\mathrm{iz} \rangle  / S_\mathrm{iz} \langle \circ \rangle$ approaching $\sim 1$.
Finally, for the positively correlated sample, we find even higher ratios, where now the fluctuation-including ionization rate $\langle S_\mathrm{iz} \rangle$ exceeds the mean-field rate $S_\mathrm{iz} \langle \circ \rangle$ by roughly a factor of 3, aligning with the turbulent rate increase observed in previous works \cite{leddy2017, thrysoe2016, umansky2024}.


\section{Conclusions}
\label{sec:conclusion}

Non-linear atomic reaction rates are essential for modeling the edge and scrape-off layer plasma.
In this paper, we have reported on how such rates are influenced by the inclusion/omission of turbulent plasma fluctuations.
Two simulations of ASDEX-Upgrade have been performed, in attached and detached plasma conditions, with low ($<50\%$) and high ($>300\%$) fluctuation amplitudes, respectively.
We have compared hydrogenic ionization, recombination, and impurity radiation rates obtained by including fluctuations, $\reftwo{\langle S \rangle}$, with rates obtained by pre-averaged, mean-field inputs, $\reftwo{ S \langle \circ \rangle}$.

While differences between the two approaches are marginal in the attached reference case ($\sim 1\%$ relative deviation), significant discrepancies arise in the detached case. 
Local to the detachment front, the fluctuation-including ionization source rate $\langle S_\mathrm{iz} \rangle$ and impurity radiation density $\langle P_\mathrm{rad}\rangle$ are lower than the corresponding mean field rates by approximately a factor of 2. 
For recombination, the fluctuation-including sink rate $\langle S_\mathrm{rc} \rangle$ increases in magnitude compared to the mean-field variant by a factor of at least 4 due to individual fluctuations (though not the bulk plasma) reaching the threshold temperatures and densities for the plasma to effectively recombine.
\reftwo{While recombination is enhanced by fluctuations reaching the ionization-recombination threshold energy $(\sim 2-3\,\mathrm{eV})$, ionization is analogously reduced.}
Reduced mean temperatures together with larger fluctuation amplitudes in the vicinity of the detachment front help enable this reduction.
\reftwo{Notably, these effects occur primarily in a highly localized region where the detachment front enters the confined edge region, while otherwise the difference between averaging orders becomes marginal.}

We further identify that the correlation of density and temperature fluctuations plays a significant role. 
Contrary to the \refone{standard blob structure\cite{Kube2019,zhang2024}} at the outboard-midplane (positive density-temperature correlation, i.e.~hot and dense), we find predominantly negatively correlated fluctuation samples in the confined region near the X-point (i.e.~cold and dense).
This holds for both simulations (with and without detachment fronts), though no effect is observed in the attached case as the temperature threshold is not reached, even when accounting for fluctuations.

To further explore the role of correlations, we have constructed two synthetic distributions with averages and fluctuation amplitudes identical to those observed at the X-point, but with decorrelated and positively correlated density-temperature pairs.
For decorrelated samples, ionization reduction no longer occurs, and the fluctuation-including and mean-field rates equalize.
For positively correlated samples, the trend reverses instead, and the fluctuation-including ionization rate becomes $\sim 3$ times higher than the mean-field rate.
This observation is consistent with previous studies on reaction rates \cite{leddy2017, thrysoe2016, umansky2024}.

\reftwo{We can thus corroborate earlier findings \cite{marandet2011, thrysoe2016, umansky2024} that ionization and recombination are significantly affected in cold, large fluctuation conditions,} now for the first time in realistic conditions well matched with a detached divertor experiment.
Although positively correlated fluctuations could, in principle, enhance the effective ionization rate, our simulations show that the plasma near the detachment front exhibits negative correlation, which instead reduces ionization.
Together with the fact that fluctuations effectively activate a recombination sink, we infer that their combined impact is an effective halving of the plasma density source.

\reftwo{While the neutral gas and impurity models in our turbulence simulations require future improvements as well, e.g.~kinetic neutrals and molecules \cite{Fantz2001,Verhaegh2024,rensink1998,uytven2020,uytven2022,holm2021,borodin2022}, we expect our findings to remain true qualitatively as they concern the fluctuation bias in the evaluation of \textit{any} reaction rates.} 
This suggests that mean-field modeling results must be interpreted with care, as the reaction rates in detached conditions are likely affected by turbulence.
\refone{Experimental measurements of divertor fluctuations\cite{Grenfell2024,Hoefler2025} could be key in confirming our findings.} 
At the same time, mean-field modellers could explore how much their results are affected if reaction rates are varied in the limits suggested by our simulations, \reftwo{e.g.~by a factor of two}.

\appendix
\reftwo{\section{The \grillix\, model}}
\label{sec:appendix}

The plasma model of \grillix\,is based on the drift-reduced Braginskii model \cite{zeiler1997, zeiler:habil99, scott:habil01, zholobenko2021a, zholobenko2024}, which is suitable to describe low-frequency turbulent dynamics $\omega \ll \Omega_i$ in the collisional limit. Here, $\Omega_i=eB/m_\mathrm{i}$ is the ion cyclotron frequency. Within the drift approximation, the ion and electron fluid velocities are decomposed into perpendicular drift motions and parallel streaming according to:
\begin{align*}
\mathbf{v}_\mathrm{e} = & \frac{\mathbf{B}\times\nabla\phi}{B^2} - \frac{\mathbf{B}\times\nabla p_\mathrm{e}}{enB^2} + v_{\parallel,\mathrm{e}}\mathbf{b}, \\
\mathbf{v}_\mathrm{i} = & \frac{\mathbf{B}\times\nabla\phi}{B^2} + \frac{\mathbf{B}\times\nabla p_\mathrm{i}}{enB^2} + v_{\parallel,\mathrm{i}}\mathbf{b} +\mathbf{v}_{\mathrm{pol},\mathrm{i}} \,,
\end{align*}
where the first term on the right-hand sides corresponds to the $\mathbf{E}\times\mathbf{B}$ drift $\mathbf{v}_\mathrm{E}$, and the second term represents the electron $\mathbf{v}_\mathrm{de}
$ and ion diamagnetic $\mathbf{v}_\mathrm{di}$ drifts, respectively. The third term represents parallel streaming along the magnetic field, with its unit vector $\mathbf{b}= \mathbf{b}_0 + \mathbf{b}_1 =  \mathbf{B}/B + \nabla\times(A_\parallel \mathbf{B}/B)/B$, which includes both the equilibrium magnetic field $\mathbf{B}$ and magnetic fluctuations \cite{zhang2024,zhang2025}. The ion polarization drift $\mathbf{v}_{\mathrm{pol},\mathrm{i}}$ is formally of higher order in the drift expansion, but must be retained as its divergence is of the same order as the divergence of the higher order drifts.

The equations evolved are the electron continuity equation under the assumption of quasi-neutrality $(n=n_\mathrm{e}=n_\mathrm{i})$, which becomes:
\begin{align}\label{eq:brag_cont}
\frac{\partial n}{\partial t} + \left(\mathbf{v}_E + v_{\parallel,\mathrm{i}}\mathbf{b}\right) \cdot\nabla n = & nC(\phi)-\frac{n}{e}C\left(p_\mathrm{e}\right)+\frac{1}{e}\nabla\cdot\left(j_\parallel \mathbf{b}\right) \notag \\ 
& +S_n \,,
\end{align}
where we introduced the the parallel current $j_\parallel=en\left(v_{\parallel,\mathrm{i}}-v_{\parallel,\mathrm{e}}\right)$ and the curvature operator $C(f)=-\nabla\cdot(\mathbf{B}/B^2\times\nabla f)$. The quasi-neutrality condition $\nabla\cdot\mathbf{j}=0$ can be expressed as
\begin{align}\label{eq:brag_vort}
\nabla\cdot\left[\frac{m_\mathrm{i}n}{B^2}\frac{d_\mathrm{i}}{dt}\left(\nabla_\perp\phi+\frac{\nabla_\perp p_\mathrm{i}}{en}\right)\right] + \frac{C(G)}{6} = & -C(p_\mathrm{e}+p_\mathrm{i}) \notag \\ 
& + \nabla\cdot\left(j_\parallel \mathbf{b}\right) + S_\Omega \,,
\end{align}
with the ion advective derivative $d_\mathrm{i}/dt = \partial_t+\left(\mathbf{v}_\mathrm{E}+v_{\parallel,\mathrm{i}}\mathbf{b} + \mathbf{v}_{\mathrm{pol},\mathrm{i}}\right)\cdot\nabla$, and the ion stress function
\begin{align*}
G=-\eta_\mathrm{flow}\left[\frac{2}{B^{3/2}}\nabla\cdot\left(v_{\parallel,\mathrm{i}}B^{3/2}\mathbf{b}\right)-\frac{C(\phi)}{2}-\frac{C(\phi)}{2en}\right] \,.
\end{align*}
To extend the model towards lower collisionality \cite{zholobenko2024}, a neoclassical correction is applied such that
\begin{align*}
{\eta}_\mathrm{flow} = \frac{\eta_{\mathrm{i}}^\mathrm{B} }{(1 + \nu_*^{-1})(1 + \epsilon^{-3/2} \nu_*^{-1})} \,,
\end{align*}
where $\eta_{\mathrm{i}}^\mathrm{B}=0.96/\nu_\mathrm{i}$ is the Braginskii ion viscosity coefficient and $\nu_* = \nu_\mathrm{i}R q_{95}/\epsilon^{3/2} \sqrt{m_{\mathrm{i}}/(2T_{\mathrm{i}})}$, with the edge safety factor $q_{95}\approx 4$ and inverse aspect ratio $\epsilon = a/R \approx 0.3$ (taking typical AUG values). 
The further equations for the plasma model are the parallel momentum equation,
\begin{align}\label{eq:brag_upar}
\frac{d_\mathrm{i} v_{\parallel,\mathrm{i}}}{dt} = -\frac{\mathbf{b}\cdot\nabla(p_\mathrm{e} + p_\mathrm{i})}{m_\mathrm{i}n} + \frac{T_\mathrm{i}}{e}C(v_{\parallel,\mathrm{i}}) - \frac{2 B^{3/2}}{3m_\mathrm{i}n} \mathbf{b}\cdot\nabla\frac{G}{B^{3/2}} + S_{v_{\parallel,\mathrm{i}}}\,,
\end{align}
and Ohm's law,
\begin{align}\label{eq:brag_ohm}
 \frac{\partial A_\parallel}{\partial t}+\frac{m_\mathrm{e}}{e}\frac{d_\mathrm{e}}{dt}\frac{j_\parallel}{en} =  \mathbf{b} \cdot\nabla\phi - \frac{\mathbf{b} \cdot\nabla p_e }{en} - \eta_\parallel j_\parallel + \frac{0.71}{e}\mathbf{b} \cdot \nabla T_\mathrm{e} \,,
\end{align}
with the electron mass $m_\mathrm{e}$, the parallel resistivity $\eta_\parallel$, and the electron advective derivative $d_\mathrm{e}/dt = \partial_t+\left(\mathbf{v}_\mathrm{E}+v_{\parallel,\mathrm{e}}\mathbf{b} \right)\cdot\nabla$. The parallel component of the perturbed electromagnetic potential is obtained from Ampère's law,
\begin{align}\label{eq:brag_apar}
\nabla_\perp^2A_\parallel = - \mu_0j_\parallel \,.
\end{align}
The electron temperature equation reads
\begin{align}\label{eq:brag_te}
    \frac{3}{2}\frac{d_\mathrm{e}}{dt}T_\mathrm{e}= &-T_\mathrm{e}\left[\frac{C(p_\mathrm{e})}{en}-  C(\phi)+\nabla\cdot \left(v_{\parallel,\mathrm{e}}\mathbf{b}\right)\right] \notag \\ 
&-\frac{5T_\mathrm{e}}{2e}C(T_\mathrm{e})+\frac{0.71T_\mathrm{e}}{en}\nabla \cdot (j_\parallel \mathbf{b}) + \frac{1}{n}\nabla\cdot ({q}_{\parallel,\mathrm{e}} \mathbf{b}) + \frac{\eta_\parallel}{n} j_\parallel^2 \notag \\
& -\frac{3m_\mathrm{e}\nu_\mathrm{e}}{m_\mathrm{i}}n(T_\mathrm{e}-T_\mathrm{i}) + S_{T_\mathrm{e}} \,,
\end{align}
and the ion temperature equation:
\begin{align}\label{eq:brag_ti}
\frac{3}{2} \frac{d_\mathrm{i} T_\mathrm{i}}{dt} = &-T_\mathrm{i}\left[\frac{C(p_\mathrm{e})}{en}- C(\phi)+\nabla\cdot \left(v_{\parallel,\mathrm{e}}\mathbf{b}\right)-\frac{j_\parallel   \mathbf{b}\cdot\nabla n}{en^2}\right] \notag \\ 
&  + \frac{G^2}{3n\eta_\mathrm{i}} +\frac{5T_\mathrm{i}}{2e}C(T_\mathrm{i})+\frac{1}{n}\nabla\cdot ({q}_{\parallel,\mathrm{i}} \mathbf{b})+\frac{3m_\mathrm{e}\nu_\mathrm{e}}{m_\mathrm{i}}n(T_\mathrm{e}-T_\mathrm{i}) \notag \\
& + S_{T_\mathrm{i}} \,.
\end{align}
The system is closed with the parallel heat fluxes given as $q_{\parallel,s} = \chi_{\parallel,s}\mathbf{b}\cdot\nabla T_s$, where $s\in\left\lbrace \mathrm{e},\mathrm{i}\right\rbrace$. For the parallel heat conductivities, we use free-streaming limited expressions \cite{zholobenko2024},
\begin{align*}
\chi_{\parallel,s} = \chi_{\parallel,s}^B\left(1+\frac{\chi_{\parallel,s}^B\sqrt{m_s}}{\alpha_s q_{95} Rn\sqrt{T_s}}\right)^{-1} \,,
\end{align*}
with $\chi_{\parallel,s}^B$ the Braginksii heat conductivities and $\alpha_s$ free parameters to be tuned to the particular case.

The \grillix\ model is a full-$f$ formulation and can therefore accommodate arbitrary fluctuation amplitudes. It is electromagnetic, incorporating inductive effects and magnetic flutter terms.

For technical and numerical reasons, a few modifications are employed to the overall model stated here. The most important of these are the inclusion of hyperviscosity terms and the neglect of the polarization velocity in the advective derivative, and the usage of a dynamical high-pass filter for magnetic fluctuations \cite{zhang2025}. Further details can be found in previous \grillix\ publications \cite{zholobenko2024, eder2025a} and in the \grillix\ code repository \cite{phoenix-public:grillix26}.

The fluid neutrals model \cite{uytven2020,zholobenko2021b,eder2025a} consists of dynamical equations for the neutrals density $N$,
\begin{align}\label{eq:neut_dens}
    \frac{\partial N}{\partial t}
    = \nabla \cdot \left[ 
        \frac{1}{m_\mathrm{i} n k_\mathrm{cx}} \nabla_\perp \left( N T_{\mathrm{N}} \right) 
        - \Gamma_\mathrm{\parallel, N} \mathbf{b}_0\right]
    + S_N \, ,
\end{align}
the neutrals parallel momentum $\Gamma_{\parallel,\mathrm{N}}$,
\begin{align}\label{eq:neut_parmom}
    \frac{\partial \Gamma_{\parallel,\mathrm{N}}}{\partial t} 
    = \nabla \cdot \left[ 
        \frac{1}{m_\mathrm{i} n k_\mathrm{cx}} \nabla_\perp \left( \Gamma_{\parallel,\mathrm{N}} T_{\mathrm{N}} \right)
        - \frac{\Gamma_{\parallel,\mathrm{N}}^2}{N} \mathbf{b}_0
    \right]
    - \frac{\mathbf{b}_0 \cdot \nabla p_\mathrm{N}}{m_\mathrm{i}} 
    + S_\Gamma \, ,
\end{align}
and the neutrals pressure $p_\mathrm{N}$,
\begin{align}\label{eq:neut_pres}
    \frac{\partial p_\mathrm{N}}{\partial t} 
    = & \nabla \cdot \left[
        \frac{5}{3} \frac{1}{m_\mathrm{i} n k_\mathrm{cx}} \nabla_\perp \left( p_\mathrm{N} T_\mathrm{N} \right)
        - \frac{5}{3} p_\mathrm{N} v_{\parallel,\mathrm{N}} \mathbf{b}_0 
    \right] \\ \notag 
    &+ \frac{2}{3} v_{\parallel,\mathrm{N}} \mathbf{b}_0 \cdot \nabla p_\mathrm{N} 
    + \frac{2}{3} \frac{p_\mathrm{N}}{n k_\mathrm{cx}} \left( \nabla_\perp v_{\parallel,\mathrm{N}} \right)^2 
    + S_p \, ,
\end{align}
where $v_{\parallel,\mathrm{N}} = \Gamma_{\parallel,\mathrm{N}}/N$ and $T_\mathrm{N}=p_\mathrm{N}/N$. 

The neutrals model is coupled to the plasma model via source terms.
These act on the right-hand sides of plasma equations \eqref{eq:brag_cont}, \eqref{eq:brag_vort}, \eqref{eq:brag_upar}, \eqref{eq:brag_te}, \eqref{eq:brag_ti},
\begin{align}
S_n &= - S_N = + k_\mathrm{iz} n N - k_\mathrm{rc} n^2 \, , \\
S_{\Omega} &= - \nabla \cdot \left[
    \frac{m n N}{e B^2} \left( k_\mathrm{iz} + k_\mathrm{cx} \right) \left( \nabla_\perp \phi + \frac{\nabla_\perp p_\mathrm{i}}{e n} \right) \right] \, , \\
S_{v_{\parallel,\mathrm{i}}} & = - \frac{1}{n} ( S_\Gamma + S_n v_{\parallel,\mathrm{i}}) = N \left( k_\mathrm{iz} + k_\mathrm{cx} \right) \left( v_{\parallel,\mathrm{N}} - v_{\parallel,\mathrm{i}} \right)\, , \\
 S_{T_\mathrm{e}} & = - \frac{2}{3} \left( W_\mathrm{iz} N + W_\mathrm{rc} n \right) 
    - T_\mathrm{e} \left( k_\mathrm{iz} N - k_\mathrm{rc} n \right) \, , \\
S_{T_\mathrm{i}} & = N \left( k_\mathrm{iz} + k_\mathrm{cx} \right) 
    \left[ \left( T_\mathrm{N} - T_\mathrm{i} \right) + \frac{m_\mathrm{i}}{3} \left( v_{\parallel,\mathrm{i}} - v_{\parallel,\mathrm{N}} \right)^2 \right] \,.
\end{align}
and on the neutrals equations \eqref{eq:neut_dens}, \eqref{eq:neut_parmom}, \eqref{eq:neut_pres},
\begin{align}    
S_N 
    = &- k_\mathrm{iz} n N + k_\mathrm{rc} n^2 \, , \label{eq:source_neutrals_density}\\
S_\Gamma
    = &- k_\mathrm{iz} n N v_\mathrm{\parallel,N} + k_\mathrm{rc} n^2 v_\mathrm{\parallel,i} + k_\mathrm{cx} n N \left( v_\mathrm{\parallel,i} - v_\mathrm{\parallel,N} \right) \, , \label{eq:source_neutrals_parmom} \\
S_p
    = &- k_\mathrm{iz} n N T_\mathrm{N}
    + k_\mathrm{rc} n^2 T_\mathrm{i} 
    + k_\mathrm{cx} n N \left( T_\mathrm{i} - T_\mathrm{N} \right) \\
    &+ \frac{m_\mathrm{i} n}{3} \left( n k_\mathrm{rc} + N k_\mathrm{cx} \right) \left( v_\mathrm{\parallel,i} - v_\mathrm{\parallel,N } \right)^2 \, , \label{eq:source_neutrals_pressure}
\end{align}
with electron cooling rate coefficients $W_\mathrm{iz}, W_\mathrm{rc}$ taken from the AMJUEL database\cite{julichdata} same as $k_\mathrm{iz}, k_\mathrm{rc}$ discussed in \ref{sec:reactions}.
The charge exchange reaction rate coefficient is $k_\mathrm{cx} = 2.93 \sigma_\mathrm{cx} \sqrt{T_\mathrm{i} / m_\mathrm{i}}$, with the charge-exchange cross-section approximated as constant\cite{helander1994}, $\sigma_\mathrm{cx} \approx 7\times10^{-19}\,\mathrm{m}^2$.

\reftwo{We stress that the usage of mono-atomic neutrals, neglecting molecules, is not strictly justified in detached conditions\cite{Fantz2001,Verhaegh2024}. Comparisons of such fluid neutrals models with kinetic ones which include molecules (EIRENE), however, tend to demonstrate a decent degree of agreement for the plasma solution\cite{rensink1998,uytven2020,uytven2022}. Nonetheless, in future we aim to extend our model by either fluid \cite{holm2021} or (hybrid) kinetic \cite{borodin2022} molecules. For the present analysis, we expect only quantitative but no qualitative differences due to molecules. Rather, we stress that since we simply analyze the fluctuation bias in the evaluation of reaction rates, this effect (at least qualitatively) naturally also extends to reactions concerning molecules and others that have not been yet explicitly considered in the present work.}


\section*{Acknowledgements}
\label{sec:acknowledgements}

The authors gratefully acknowledge Christoph Pitzal, Barnabás Csillag, Chris Hill and Yu-Chih Liang for fruitful discussions.
This work has been carried out within the framework of the EUROfusion Consortium, funded by the European Union via the Euratom Research and Training Programme (Grant Agreement No 101052200 – EUROfusion). 
Views and opinions expressed are those of the author(s) only and do not necessarily reflect those of the European Union or the European Commission. 
Neither the European Union nor the European Commission can be held responsible for them. 
The simulations shown in this work were performed on the national supercomputer HPE Apollo (Hawk) at the High Performance Computing Center Stuttgart (HLRS) under the grant number GRILLIX/44281, and the EUROfusion High-Performance Computers Marconi-Fusion and Pitagora under the TSVV3 project.


\section*{Declaration of generative AI and AI-assisted technologies in the writing process}

During the preparation of this work, the author(s) used DeepL Write
and ChatGPT to generate wording suggestions. After using this tool/service, the author(s) reviewed and edited the content as needed and take(s) full responsibility for the content of the publication.


\section*{Financial disclosure}

None reported.


\section*{Conflict of interest}

The authors declare no potential conflicts of interest.




\section*{Bibliography}

\bibliographystyle{ieeetr}
\bibliography{bibliography.bib}

\end{document}